\documentclass[10pt,letterpaper]{article}
\usepackage[top=0.85in,left=2.75in,footskip=0.75in,marginparwidth=2in]{geometry}

\usepackage[utf8]{inputenc}

\usepackage{cite}

\usepackage{nameref}

\usepackage[colorlinks=true, citecolor=blue, linkcolor=blue, urlcolor=blue]{hyperref}

\usepackage{pdfpages} 

\usepackage[right]{lineno}

\usepackage{microtype}
\DisableLigatures[f]{encoding = *, family = * }

\raggedright
\usepackage{setspace} 
\usepackage{changepage}

\usepackage[aboveskip=1pt,labelfont=bf,labelsep=period,singlelinecheck=off]{caption}

\makeatletter
\renewcommand{\@biblabel}[1]{\quad#1.}
\makeatother

\usepackage{lastpage,fancyhdr,graphicx}
\usepackage{epstopdf}
\fancyheadoffset[L]{2.25in}
\fancyfootoffset[L]{2.25in}

\usepackage{color}

\definecolor{Gray}{gray}{.25}

\usepackage{graphicx}

\usepackage{sidecap}

\usepackage[T1]{fontenc} 
\usepackage{helvet}      

\usepackage{amsmath}
\usepackage{sfmath} 

\reversemarginpar 

\begin{document}
\vspace*{0.35in}

{\Large
\textbf\newline{Beyond Expertise: Stable Individual Differences in Predictive Eye–Hand Coordination}
}
\newline
\bigskip\newline
\textbf{Emiko Shishido}
    \footnote{ORCID ID: \href{https://orcid.org/0000-0002-2474-1343}
    {https://orcid.org/0000-0002-2474-1343}
    \\
    \indent Researchmap: \href{https://researchmap.jp/emikoshishido?lang=en}{https://researchmap.jp/emikoshishido?lang=en}}*
\\
\noindent National Institute for Physiological Sciences, Okazaki, 444-8585, Japan
    \footnote{National Institute for Physiological Sciences: \href{https://www.nips.ac.jp/eng/}{https://www.nips.ac.jp/eng/}}
\\
\noindent * Corresponding author:\texttt{emikosh[at mark]nips.ac.jp}

\marginpar{
\begin{flushleft}
\vspace{1.2 cm} 
\color{Gray} 
\begin{singlespace}
\textbf{Cite as: }\\
Shishido, E. (2026). Beyond Expertise: Stable Individual Differences in Predictive Eye–Hand Coordination. arXiv:2602.07816. \href{https://arxiv.org/abs/2602.07816}{https://arxiv.org/abs/2602.07816}
\end{singlespace}
\end{flushleft}
}

\section*{Abstract}
Human eye–hand coordination relies on internal forward models that predict future states and compensate for sensory delays. During line tracing, the gaze typically leads the hand through predictive saccades, yet the extent to which this predictive window reflects expertise or intrinsic individual traits remains unclear. 
\smallskip\newline
\indent In this study, I examined eye–hand coordination in professional calligraphers and non experts performing a controlled line tracing task. The temporal coupling between saccade distance (SD) and pen speed (PS) revealed substantial interpersonal variability: SD–PS peak times ranged from approximately -50 to 400 ms, forming stable, participant specific predictive windows that were consistent across trials. 
\smallskip\newline
\indent These predictive windows closely matched each individual’s pen catch up time, indicating that the oculomotor system stabilizes fixation in anticipation of the hand’s future velocity rather than relying on reactive pursuit. Neither the spatial indices (mean gaze–pen distance, mean saccade distance) nor the temporal index (SD–PS peak time) differed between calligraphers and non calligraphers, and none of these predictive parameters correlated with tracing accuracy. 
\smallskip\newline
\indent These findings suggest that diverse predictive strategies can achieve equivalent performance, consistent with the minimum intervention principle of optimal feedback control. Together, the results indicate that predictive timing in eye–hand coordination reflects a stable, idiosyncratic Predictive Protocol shaped by individual neuromotor constraints rather than by expertise or training history.


\section*{Introduction}
Eye–hand coordination is a fundamental component of human interaction with the environment, and prediction is central to this process. In voluntary actions such as reaching or object manipulation, the eyes typically move ahead of the hand. Classic studies have shown that the gaze fixates on a target object well before the hand arrives, often by nearly one second \cite{RN1, RN2}. Saccadic eye movements are tightly linked to limb movements, and this coupling provides a temporal advantage: saccades become faster and more efficient when accompanied by coordinated arm movements \cite{RN3}. These findings indicate that the brain relies not only on sensory input but also on internal forward models that predict future states and guide ongoing motor behavior.
\bigskip
\newline
Forward models are known to operate robustly and often independently of conscious strategies. For example, implicit motor plans can override explicit instructions even when the explicit strategy is successful \cite{RN4}, suggesting that predictive mechanisms prioritize the minimization of sensory prediction errors over task level goals. Although forward models have been extensively studied, the degree to which individuals differ in how far ahead they pre calculate future actions remains poorly understood.
\bigskip
\newline
Line tracing provides a useful framework for examining predictive eye–hand coordination (\hyperlink{Supplementary video}{Supplementary video}). During tracing, the gaze typically alternates between saccades and fixations, leading the hand along the path \cite{RN5}. A saccade shifts the gaze forward, and the pen subsequently catches up during the following fixation. While the neural mechanisms underlying this dual system behavior remain unclear, it is plausible that individuals adopt stable, idiosyncratic strategies—a Predictive Protocol—to coordinate gaze and hand movements in a way that minimizes errors. These predictive mechanisms operate continuously during movement execution and are largely outside conscious awareness.
\bigskip
\newline
In earlier work, we developed a controlled line tracing task to extract the features of gaze and hand movements, and part of the works was published in the Japanese journal IEICE Tech. Rep., as a technical report \cite{RN6}. That study focused on group level differences between expert calligraphers and non experts. The calligraphers were practitioners of traditional Japanese brush writing (shodō), a form of large‑scale brush‑and‑ink distinct from Western pen‑based calligraphy. However, visual inspection of the raw data suggested that each participant exhibited a consistent coordination pattern across trials, hinting at substantial individual variability that was not captured by group comparisons.
\bigskip
\newline
The present study analyzes the extended dataset from the previous work, with a focus on individual differences in predictive timing. Specifically, I examined whether each participant exhibited a stable predictive window between saccadic prediction and hand movement, and whether this window varied systematically across individuals. I further tested whether these temporal and spatial strategies were related to tracing accuracy. Preliminary observations indicated that accuracy remained largely independent of an individual’s Predictive Protocol, suggesting that the central nervous system flexibly adjusts control parameters to achieve consistent performance.
\bigskip
\newline
Taken together, this study aims to characterize the diversity of predictive timing in eye–hand coordination at the individual level and to evaluate whether these differences reflect intrinsic motor control strategies rather than expertise dependent adaptations. The findings are interpreted within the framework of optimal feedback control \cite{RN7}, particularly the minimum‑intervention principle, which predicts that variability is tolerated in dimensions that do not compromise task success.
\bigskip
\newline
\section*{Materials and Methods}
\subsection*{Participants}
Seventeen right‑handed adults (age 20–56 years, mean ± standard deviation: 40 ± 13 years) participated in the study. Seven were professional calligraphers trained in traditional Japanese brush writing (shodō) and had teaching experience. The remaining ten participants had no formal calligraphy training. All participants had normal or corrected to normal vision and provided written informed consent. The study was approved by the ethics committee of the National Institute for Physiological Sciences (EC01 039).

\subsection*{Apparatus and Experimental Setup}
Eye movements were recorded using a Tobii T60 XL eye tracker (60 Hz). Hand movements were recorded using a capacitive transparent touch panel (Magic Touch, KEYTEC, USA) placed over the T60 XL display (\hyperlink{Figure 1}{Figure 1A}). Participants traced lines on the screen using a felt tip pen while seated with head and arm support to minimize motion artifacts.
\bigskip
\newline
Pen position data were recorded on a separate Windows PC. To ensure precise temporal alignment between the eye tracker and the pen input system, both devices were synchronized using an in house circuit that generated a square wave trigger signal. Latency and timestamp consistency were validated using a high speed camera, revealing minimal temporal variance (standard deviation: 6 ms). Both systems sampled at 60 Hz, providing sufficient resolution to analyze the temporal relationship between saccadic timing and pen movement.

\subsection*{Task and Stimuli}
Participants performed a line tracing task under two speed conditions:
\smallskip\newline
\indent•	\textbf{Low speed}: target line visible for 4.5 s
    \smallskip\newline
\indent•	\textbf{High speed}: target line visible for 3.0 s
    \smallskip\newline
\noindent Each condition was presented in a separate block. Participants sat approximately 50 cm from the display and were instructed to “trace the line as accurately as possible” within the allotted time. They wore a black glove on the right hand to minimize infrared interference with the eye tracker and rested their elbow on the table (\hyperlink{Figure 1}{Figure 1A}).
\bigskip
\newline
Each trial involved tracing one of 30 pre generated lines from left to right. The lines were based on the minimum jerk model (MJM) to ensure smooth, human like trajectories (\hyperlink{Figure 1}{Figure 1B}). Three types of trajectories were included:
\smallskip\newline
\indent•	no via points (6 lines)\smallskip\newline
\indent•	single via point (12 lines)\smallskip\newline
\indent•	double via points (12 lines)\smallskip\newline
\noindent All trajectories were normalized to a total path length of 250 mm. One of the 30 lines was randomly selected for each trial.

\subsection*{Data Processing and Analysis}
Data were analyzed using MATLAB. Trials with gaze acquisition rates below 80\% were excluded. Gaze and pen positions were mapped to the nearest point on the target trajectory using least squares fitting. \textbf{Pen speed (PS)} was computed from smoothed position data using a moving average. Movement onset was defined as the first time PS exceeded 20 mm/s (\hyperlink{Figure 1}{Figure 1C and D}).
\bigskip
\newline
Saccades were identified using a velocity threshold of 31.8°/s. Fixations were identified using a velocity and duration. The \textbf{gaze–pen distance (GP)} was defined as the spatial lead of the gaze relative to the pen along the tracing path. For each saccade, the saccade distance (SD) was computed as the displacement along the path at the moment of saccade onset (\hyperlink{Figure 1}{Figure 1C and D}). Individual mean values of GP and SD were denoted as \textbf{mGP} and \textbf{mSD}.

\subsection*{Temporal Correlation Between Saccade Distance and Pen Speed}
For each saccade event across all trials, pen speed (PS) was sampled at discrete time points before and after saccade onset at 1/60 s intervals ($\approx$ 16.7 ms). Time lags ranged from -215.8 ms to +431.6 ms relative to saccade onset. At each lag, the correlation between SD and PS was computed using robust linear regression (\hyperlink{Figure 2}{Figure 2A}). The time lag yielding the highest correlation was defined as the \textbf{SD–PS peak time}, representing the individual’s predictive window (\hyperlink{Figure 2}{Figure 2B}).
\bigskip
\newline
To relate the SD–PS peak time to spatial dynamics, the \textbf{pen catch up time} was defined as the interval between saccade onset and the moment when the pen reached the spatial position of the preceding fixation. Positive values indicate that the gaze leads the pen; negative values indicate that the pen precedes the gaze.

\subsection*{Tracing Accuracy}
Tracing accuracy was quantified as the mean absolute deviation between the pen position and the centerline of the target trajectory at each time point. The average deviation across all valid trials was used as each participant’s accuracy metric (\textbf{mError}).

\section*{Results}
\subsection*{Individual Differences in Gaze–Pen Coordination}
Participants maintained consistent drawing speeds across conditions (high speed: 180.6 ± 23.97 mm/s; low speed: 121.5 ± 19.13 mm/s). Under these uniform constraints, pronounced individual differences emerged in gaze–hand coordination (\hyperlink{Figure 2}{Figure 2}). Visual inspection revealed stable, participant specific patterns across trials (\hyperlink{Supplementary Figure S1}{Supplementary Figure S1}).
\bigskip
\newline
The mean gaze–pen distance (mGP) across participants was 16.9 ± 16.0 mm  (\hyperlink{Supplementary Figure S2}{Supplementary Figure S2}). Most participants (15/17) showed positive mGP values, indicating a gaze lead strategy, while two participants showed negative mGP values. The mean saccade distance (mSD) was 43.40 ± 8.43 mm (\hyperlink{Supplementary Figure S2}{Supplementary Figure S2}). mGP and mSD were positively correlated (R = 0.66, P = 0.003).

\subsection*{Predictive Saccades and Pen Speed}
All participants showed a significant positive correlation between SD and subsequent PS (\hyperlink{Figure 4}{Figures 4 and }\hyperlink{Figure 5}{5}). However, the SD–PS peak time, which represents the predictive window, varied widely across individuals, ranging from approximately -50 to 400 ms (\hyperlink{Figure 4}{Figures 4 and }\hyperlink{Figure 5}{5}). Fourteen participants (82\%) showed peak times > 0 ms, indicating prediction of future hand speed; three showed peak times < 0 ms.
\bigskip
\newline
The mean peak correlation coefficient was R = 0.59 ± 0.10. No significant differences were found between calligraphers and non calligraphers (P > 0.05).

\subsection*{Pen Catch Up Time and Predictive Timing}
The mean pen catch up time varied across participants and was strongly correlated with the SD–PS peak time (R = 0.80, P = 4.3 × $10^{-5}$; \hyperlink{Figure 6}{Figures  6 and }\hyperlink{Figure 7}{7}). This indicates that each individual’s predictive window is tightly linked to the temporal dynamics of how the hand approaches the gaze position.

\subsection*{Predictive Protocols and Tracing Accuracy}
Tracing accuracy (mean error: 1.54 ± 0.22 mm) did not correlate with mGP (R = 0.13, P = 0.61), mSD (R = 0.05, P = 0.84), or SD–PS peak time (R = 0.001, P = 1.00; \hyperlink{Figure 8}{Figure 8}). Despite diverse Predictive Protocols, all participants achieved comparable accuracy.
\bigskip
\newline
Calligraphers and non calligraphers did not differ in tracing accuracy (P > 0.05).

\section*{Discussion} 
The present study investigated predictive eye–hand coordination during a controlled line tracing task, focusing on stable individual differences rather than group level effects. Despite identical temporal and spatial constraints, participants did not converge on a single predictive strategy. Instead, each individual exhibited a consistent and highly personalized Predictive Protocol that governed the timing and spatial organization of gaze–hand interactions.
\bigskip
\newline
\subsection*{Temporal diversity in predictive saccades}
A central finding is the substantial interpersonal variability in the temporal coupling between saccade distance (SD) and pen speed (PS). Although all participants showed a significant positive correlation between SD and subsequent PS, the SD–PS peak time, which represents the predictive window, varied widely across individuals, ranging from approximately -50 to 400 ms. This broad distribution suggests that each participant operates with a distinct predictive window. Some individuals relied on short range prediction tightly linked to immediate hand control, whereas others maintained a long range window, anticipating hand speed nearly 400 ms in advance. Because these patterns were consistent across trials and speed conditions, they likely reflect stable characteristics of each participant’s motor control system.
\bigskip
\newline
Importantly, the SD–PS peak time closely matched each participant’s pen catch up time. Although the proximity of the pen to the gaze might typically evoke smooth pursuit, the present results suggest a more proactive mechanism. The oculomotor system appears to anticipate the velocity at which the pen will cross the gaze position at the moment of saccade programming, rather than relying on reactive pursuit. This interpretation aligns with evidence that smooth pursuit can be actively suppressed when it interferes with other motor demands. Electrical stimulation of the frontal eye field of non‑human primates can decelerate pursuit movements \cite{RN8}. In humans, Borot et al. reported concurrent upper limb movements can degrade pursuit performance \cite{RN9}. In a complex task such as line tracing, the brain may therefore prioritize a predict and fixate strategy to maintain manual precision.

\subsection*{Innate Predictive Protocols vs. acquired skill}
No significant differences were observed between professional calligraphers and non experts in SD–PS peak time, pen catch up time, or spatial gaze metrics. This suggests that the depth of the predictive window is not shaped by calligraphic training but instead reflects a more fundamental individual characteristic.
\bigskip
\newline
Although the number of expert calligraphers was relatively small (n = 7), the primary aim of this study was not to detect group level differences but to characterize stable individual Predictive Protocols. Effect sizes for expertise related differences were negligible across all predictive metrics, making it unlikely that the absence of group differences is due solely to limited statistical power. Moreover, the variance attributable to individual differences far exceeded that attributable to expertise, indicating that predictive timing is dominated by person specific factors rather than training history. Future studies with larger expert samples may further examine whether subtle expertise related modulations exist, but the present findings strongly support the interpretation that the predictive window reflects an intrinsic individual characteristic.
\bigskip
\newline
This perspective suggests that calligraphy training enhances execution level precision without altering the temporal window of the internal forward model. The capacity for “future thinking”—the ability to simulate upcoming states before they occur—may therefore represent a pervasive trait that influences not only motor coordination but also broader cognitive processes such as risk management. The wide variance in predictive windows observed here may be the behavioral manifestation of an individual’s Predictive Protocol, determining how much cognitive and neural resource is allocated to anticipating future states.

\subsection*{Predictive Protocols in relation to prior work}
The present findings extend previous work by demonstrating that individual differences in predictive timing can be detected even in a small sample performing a naturalistic tracing task. Prior studies have emphasized group level comparisons \cite{RN6} but the current approach highlights the fine grained ways in which predictive mechanisms unfold during real time skilled actions.
\bigskip
\newline
The stability of the predictive window across skill levels aligns with the view that predictive adaptation is a low level, autonomous process. Implicit motor adaptation is driven by sensory prediction errors and is largely resistant to conscious strategies \cite{RN4}. This supports the hypothesis that the predictive window represents an innate Predictive Protocol rather than an acquired skill.
\bigskip
\newline
The observed individual differences may also reflect personal baselines for sensory integration. According to optimal integration theory \cite{RN10}, the brain combines internal predictions with sensory feedback by weighting each according to its reliability. In this context, the Predictive Protocol may represent an individual’s characteristic weighting between forward model predictions and delayed visual feedback during high precision tasks.
\bigskip
\newline
Optimal feedback control provides a useful theoretical framework for interpreting these findings \cite{RN7}. In particular, the minimum‑intervention principle predicts that variability is tolerated in dimensions that do not compromise task success. This perspective aligns with evidence that internal monitoring of limb dynamics is continuous and robust: variations in arm dynamics require the brain to track hand position using predictive models \cite{RN11}, and perturbations to saccade trajectories are corrected within the same movement \cite{RN11}.
\bigskip
\newline
In this view, the wide interpersonal variability observed in the predictive window is not a deviation from optimal control but an expression of it. Each participant appears to adopt a personalized balance between prediction and feedback, selecting control policies that stabilize task‑relevant dimensions while allowing flexibility elsewhere. Such individualized solutions are precisely what optimal feedback control predicts when multiple strategies can achieve equivalent performance. The Predictive Protocol identified here may therefore represent a behavioral signature of how each person’s motor system resolves the optimal‑control trade‑off under shared task constraints.

\subsection*{Conclusion}
This study examined predictive eye–hand coordination during a controlled line tracing task and revealed that individuals rely on stable, person specific Predictive Protocols rather than a universal predictive mechanism. Despite identical task constraints, participants exhibited distinct predictive windows in which saccade distance predicted hand speed, ranging from approximately -50 to 400 ms. These predictive windows closely matched each participant’s pen catch up time, indicating that the oculomotor system stabilizes fixation in anticipation of the hand’s future velocity.
\bigskip
\newline
Importantly, these predictive parameters did not differ between professional calligraphers and non experts, nor were they associated with tracing accuracy. The absence of expertise related differences, combined with negligible effect sizes, indicates that the Predictive Protocol reflects an intrinsic characteristic of each individual’s motor control system rather than a skill acquired through training. This interpretation aligns with the minimum intervention principle of optimal feedback control, which posits that variability is tolerated as long as task goals are preserved.
\bigskip
\newline
Together, these findings suggest that predictive timing in eye–hand coordination is shaped by individual neuromotor constraints and sensory integration strategies. Understanding these idiosyncratic Predictive Protocols may provide a foundation for future work examining how internal models are calibrated across different tasks, sensory environments, and developmental or clinical populations. In this sense, the diversity of Predictive Protocols observed here may reflect individualized solutions within an optimal‑feedback‑control framework, where multiple predictive strategies can achieve equivalent task performance.

\subsection*{Limitations}
This study has several limitations. First, the analysis primarily treated eye movements during tracing as a sequence of saccades and fixations. However, as indicated in \hyperlink{Figure 1}{Figure 1C}, some participants—particularly those with shorter gaze–pen distances—exhibited a greater proportion of smooth pursuit movements. Because the temporal and spatial dynamics of pursuit differ from those of discrete saccades, the presence of pursuit may introduce bias into the SD–PS correlation analysis. Although the present study focused on saccadic prediction as a marker of the Predictive Protocol, future work should examine how smooth pursuit contributes to predictive control and whether it modulates the observed individual variability.
\bigskip
\newline
Second, the sample size of expert calligraphers was relatively small (n = 7). While the primary aim was to characterize individual Predictive Protocols rather than detect group level differences, and effect sizes for expertise related differences were negligible, larger samples of experts would allow for a more detailed examination of potential training related modulations.
\bigskip
\newline
Finally, the tracing task involved simplified, two dimensional trajectories under controlled timing constraints. Although this design enabled precise quantification of gaze–hand dynamics, it may not fully capture the complexity of naturalistic handwriting or drawing. Future studies incorporating more ecologically valid tasks, higher sampling rates, or three dimensional movements may provide additional insight into how Predictive Protocols generalize across contexts. 

\section*{Supplementary Materials}
\subsection*{Supplementary Video}
\hypertarget{Supplementary video}{
A video demonstration of the task reconstructed from data is available in the ancillary files. For a quick preview, the video can also be viewed online as YouTube video. The video is also available in the ancillary files of this submission.
\newline
\indent\href{https://youtu.be/wQ4QFoVJDk8}{•	Movie 1 } (\href{https://youtu.be/wQ4QFoVJDk8}{https://youtu.be/wQ4QFoVJDk8})
\newline
\indent\href{https://youtu.be/UoWoDqWrgDw}{•	Movie 2 } (\href{https://youtu.be/UoWoDqWrgDw}{https://youtu.be/UoWoDqWrgDw})
}

\subsection*{Supplementary Figures S1 and S2}
\hypertarget{Supplementary Figure S1}{Supplementary Figure S1} and \hypertarget{Supplementary Figure S2}{S2} are provided in a separate supplementary PDF (“supplementary.pdf”) accompanying this submission.

\subsection*{Receiving Updates on Supplementary Materials}
Supplementary materials such as detailed methods, code, and data will be made available at \href{https://github.com/emiko-sh/eye-hand-coordination-2026}{GitHub} (\href{https://github.com/emiko-sh/eye-hand-coordination-2026}{https://github.com/emiko-sh/eye-hand-coordination-2026}).
GitHub users can 'Watch' the repository to be notified of updates. Users can open the “Watch” pull‑down menu on the top-left corner of the repository, choose “Custom”, then enable “Releases” to receive notification of new release of materials.

\section*{Acknowledgments}
The author gratefully acknowledges Norihiro Sadato, Tetsuya Yamamoto, and Masaki Fukunaga of the National Institute for Physiological Sciences; Norio Ozaki and Seiko Miyata of the Nagoya University Graduate School of Medicine; and Naohiro Fukumura of the Toyohashi University of Technology for their insightful discussions and for providing an excellent research environment.

This work was supported by JSPS KAKENHI Grant Number 24K10723.


\bibliography{library}

@article{RN9,
   author = {Borot, L. and Ogden, R. and Bennett, S. J.},
   title = {Prefrontal cortex activity and functional organisation in dual-task ocular pursuit is affected by concurrent upper limb movement},
   journal = {Sci Rep},
   volume = {14},
   number = {1},
   pages = {9996},
   ISSN = {2045-2322 (Electronic)
2045-2322 (Linking)},
   DOI = {10.1038/s41598-024-57012-2},
   url = {https://www.ncbi.nlm.nih.gov/pubmed/38693184},
   year = {2024},
   type = {Journal Article}
}

@article{RN2,
   author = {Flanagan, J. R. and Vetter, P. and Johansson, R. S. and Wolpert, D. M.},
   title = {Prediction precedes control in motor learning},
   journal = {Curr Biol},
   volume = {13},
   number = {2},
   pages = {146–50},
   ISSN = {0960-9822 (Print)
0960-9822 (Linking)},
   url = {http://www.ncbi.nlm.nih.gov/pubmed/12546789
http://ac.els-cdn.com/S0960982203000071/1-s2.0-S0960982203000071-main.pdf?_tid=16404bf0-b320-11e3-879c-00000aacb35d&acdnat=1395643794_8bbd44024cd5a19f78df203168e6d0b9},
   year = {2003},
   type = {Journal Article}
}

@article{RN5,
   author = {Gowen, E. and Miall, R. C.},
   title = {Eye-hand interactions in tracing and drawing tasks},
   journal = {Hum Mov Sci},
   volume = {25},
   number = {4-5},
   pages = {568–85},
   ISSN = {0167-9457 (Print)
0167-9457 (Linking)},
   DOI = {10.1016/j.humov.2006.06.005},
   url = {https://www.ncbi.nlm.nih.gov/pubmed/16891021},
   year = {2006},
   type = {Journal Article}
}

@article{RN8,
   author = {Izawa, Y. and Suzuki, H. and Shinoda, Y.},
   title = {Suppression of smooth pursuit eye movements induced by electrical stimulation of the monkey frontal eye field},
   journal = {J Neurophysiol},
   volume = {106},
   number = {5},
   pages = {2675–87},
   ISSN = {1522-1598 (Electronic)
0022-3077 (Linking)},
   DOI = {10.1152/jn.00182.2011},
   url = {https://www.ncbi.nlm.nih.gov/pubmed/21849604},
   year = {2011},
   type = {Journal Article}
}

@article{RN1,
   author = {Johansson, R. S. and Westling, G. and Backstrom, A. and Flanagan, J. R.},
   title = {Eye-hand coordination in object manipulation},
   journal = {J Neurosci},
   volume = {21},
   number = {17},
   pages = {6917–32},
   ISSN = {1529-2401 (Electronic)
0270-6474 (Print)
0270-6474 (Linking)},
   DOI = {10.1523/JNEUROSCI.21-17-06917.2001},
   url = {https://www.ncbi.nlm.nih.gov/pubmed/11517279},
   year = {2001},
   type = {Journal Article}
}

@article{RN4,
   author = {Mazzoni, P. and Krakauer, J. W.},
   title = {An implicit plan overrides an explicit strategy during visuomotor adaptation},
   journal = {J Neurosci},
   volume = {26},
   number = {14},
   pages = {3642–5},
   ISSN = {1529-2401 (Electronic)
0270-6474 (Print)
0270-6474 (Linking)},
   DOI = {10.1523/JNEUROSCI.5317-05.2006},
   url = {https://www.ncbi.nlm.nih.gov/pubmed/16597717},
   year = {2006},
   type = {Journal Article}
}

@article{RN11,
   author = {Nanayakkara, T. and Shadmehr, R.},
   title = {Saccade adaptation in response to altered arm dynamics},
   journal = {J Neurophysiol},
   volume = {90},
   number = {6},
   pages = {4016–21},
   ISSN = {0022-3077 (Print)
0022-3077 (Linking)},
   DOI = {10.1152/jn.00430.2003},
   url = {https://www.ncbi.nlm.nih.gov/pubmed/14665687},
   year = {2003},
   type = {Journal Article}
}

@article{RN3,
   author = {Snyder, L. H. and Calton, J. L. and Dickinson, A. R. and Lawrence, B. M.},
   title = {Eye-hand coordination: saccades are faster when accompanied by a coordinated arm movement},
   journal = {J Neurophysiol},
   volume = {87},
   number = {5},
   pages = {2279–86},
   ISSN = {0022-3077 (Print)
0022-3077 (Linking)},
   DOI = {10.1152/jn.00854.2001},
   url = {https://www.ncbi.nlm.nih.gov/pubmed/11976367},
   year = {2002},
   type = {Journal Article}
}

@article{RN7,
   author = {Todorov, E. and Jordan, M. I.},
   title = {Optimal feedback control as a theory of motor coordination},
   journal = {Nat Neurosci},
   volume = {5},
   number = {11},
   pages = {1226–35},
   ISSN = {1097-6256 (Print)
1097-6256 (Linking)},
   DOI = {10.1038/nn963},
   url = {https://www.ncbi.nlm.nih.gov/pubmed/12404008},
   year = {2002},
   type = {Journal Article}
}

@article{RN10,
   author = {Vaziri, S. and Diedrichsen, J. and Shadmehr, R.},
   title = {Why does the brain predict sensory consequences of oculomotor commands? Optimal integration of the predicted and the actual sensory feedback},
   journal = {J Neurosci},
   volume = {26},
   number = {16},
   pages = {4188–97},
   ISSN = {1529-2401 (Electronic)
0270-6474 (Print)
0270-6474 (Linking)},
   DOI = {10.1523/JNEUROSCI.4747-05.2006},
   url = {https://www.ncbi.nlm.nih.gov/pubmed/16624939},
   year = {2006},
   type = {Journal Article}
}

@article{RN6,
   author = {Yamasaki, Kenshiro and Itoh, Tatsuki and Itoh, Yoshikuni and Okazaki, Shuntaro and Sadato, Norihiro and Imoto, Keiji and Shishido, Emiko and Fukumura, Naohiro},
   title = {Feature extraction of eye-hand coordination in tracing tasks of calligraphers},
   journal = {IEICE Tech. Rep.},
   volume = {114 (515) NC2014-123},
   pages = {313–318},
   year = {2015},
   type = {Journal Article}
}

\bibliographystyle{apalike}

\vspace{0cm} 
\begin{adjustwidth}{-2in}{0in}
\begin{center}
\includegraphics[width=0.72\linewidth]{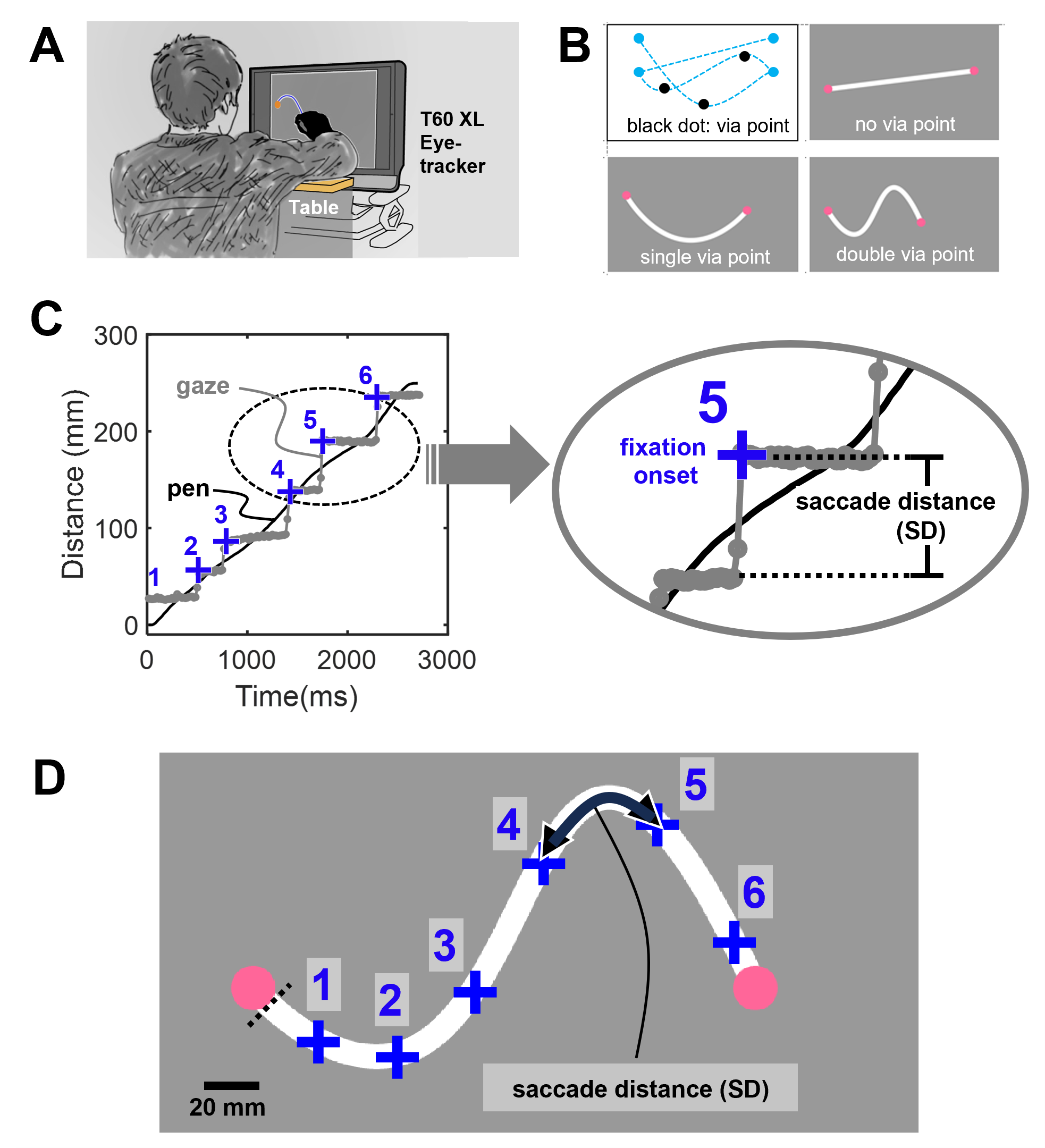} 
\end{center}
\justify 
\color{Gray}
\hypertarget{Figure 1}{\textbf{Figure 1. Experimental setup and task.}}
\newline
\textbf{A.} Schematic illustration of the experimental setup showing the synchronized eye tracking and touch panel systems. \textbf{B.} Examples of the 30 minimum jerk model (MJM) target lines used in the tracing task. \textbf{C.} Representative time series plot of gaze (gray) and pen (black) positions from a single trial. The horizontal axis (Time) is aligned to the onset of pen movement (pen speed > 20 mm/s). The vertical axis (Distance) represents the path length from the start point of the target line. Blue plus markers (+) and numbers indicate fixation onsets. Horizontal brackets indicate the saccade distance (SD) for each predictive step. \textbf{D.} Spatial organization of eye movements relative to the target. Markers and line colors match those in panel C. The dotted horizontal line indicates the threshold used to define pen movement onset.

\begin{center}
\includegraphics[width=0.9\linewidth]{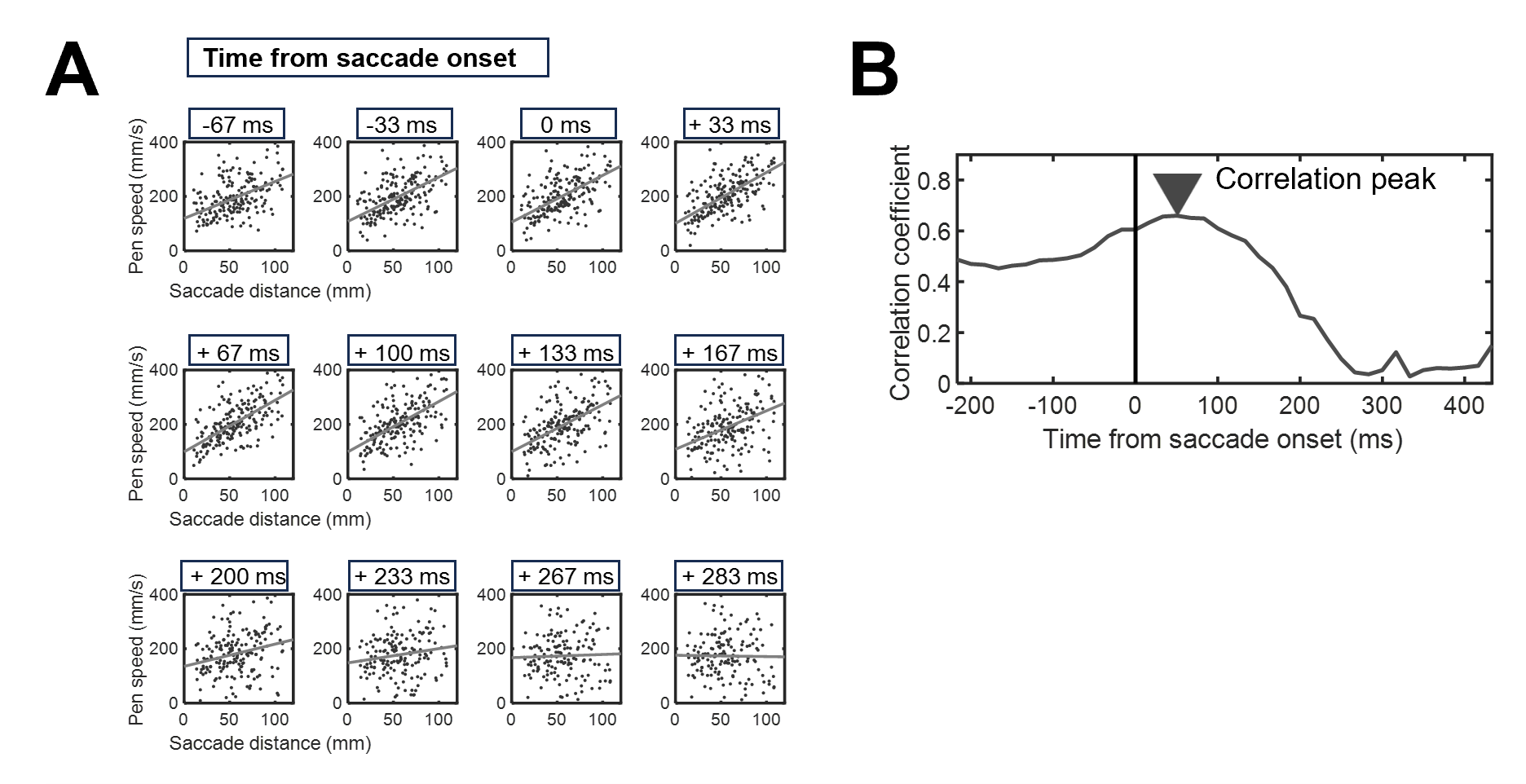} 
\end{center}
\justify 
\color{Gray}
\hypertarget{Figure 2}{\textbf{Figure 2. Temporal correlation between saccade distance (SD) and pen speed (PS).}}
\newline
\textbf{A.} Relationship between SD and PS on a saccade by saccade basis. Each dot represents an individual saccade. The solid line shows the robust linear regression. \textbf{B.} Cross correlation profile between SD and PS. The correlation coefficient (R) is plotted as a function of time lag relative to saccade onset. Positive lags indicate that SD correlates with future PS; negative lags indicate correlation with past PS. The vertical dashed line marks saccade onset (t = 0). The filled inverted triangle indicates the SD–PS peak time, representing the timing at which SD and PS exhibit the strongest temporal coupling.
\bigskip\newline

\begin{center}
\includegraphics[width=0.8\linewidth]{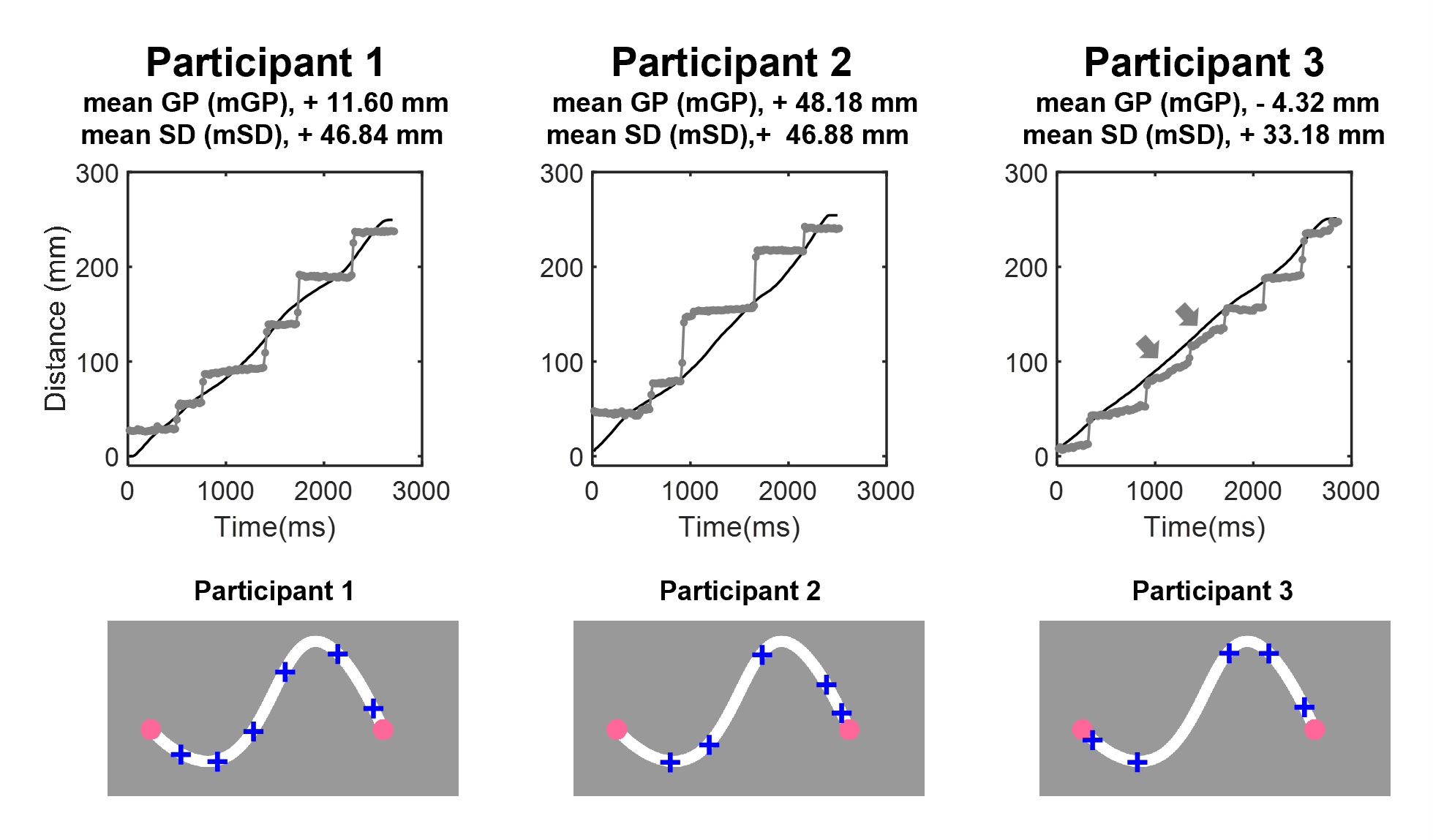} 
\end{center}
\justify 
\color{Gray}
\hypertarget{Figure 3}{\textbf{Figure 3. Individual diversity in predictive strategies.}}
\newline
Representative time series plots from three participants (Participants 2, 3, and 5). Axes and markers follow the definitions in \textbf{Figure 1C}. Although all participants accurately tracked the target, their spatial strategies varied substantially. Values for mean gaze–pen distance (mGP) and mean saccade distance (mSD) are shown in each panel. Participant 2 exhibits a large gaze lead strategy (large mGP), whereas Participant 3 shows a much smaller gaze lead, and on average the gaze position lags behind the pen (negative mGP). Arrows in Participant 3 highlight instances of target derived smooth pursuit movements as the pen passes near the gaze position. All traces are from the low speed condition.

\begin{center}
\includegraphics[width=0.8\linewidth]{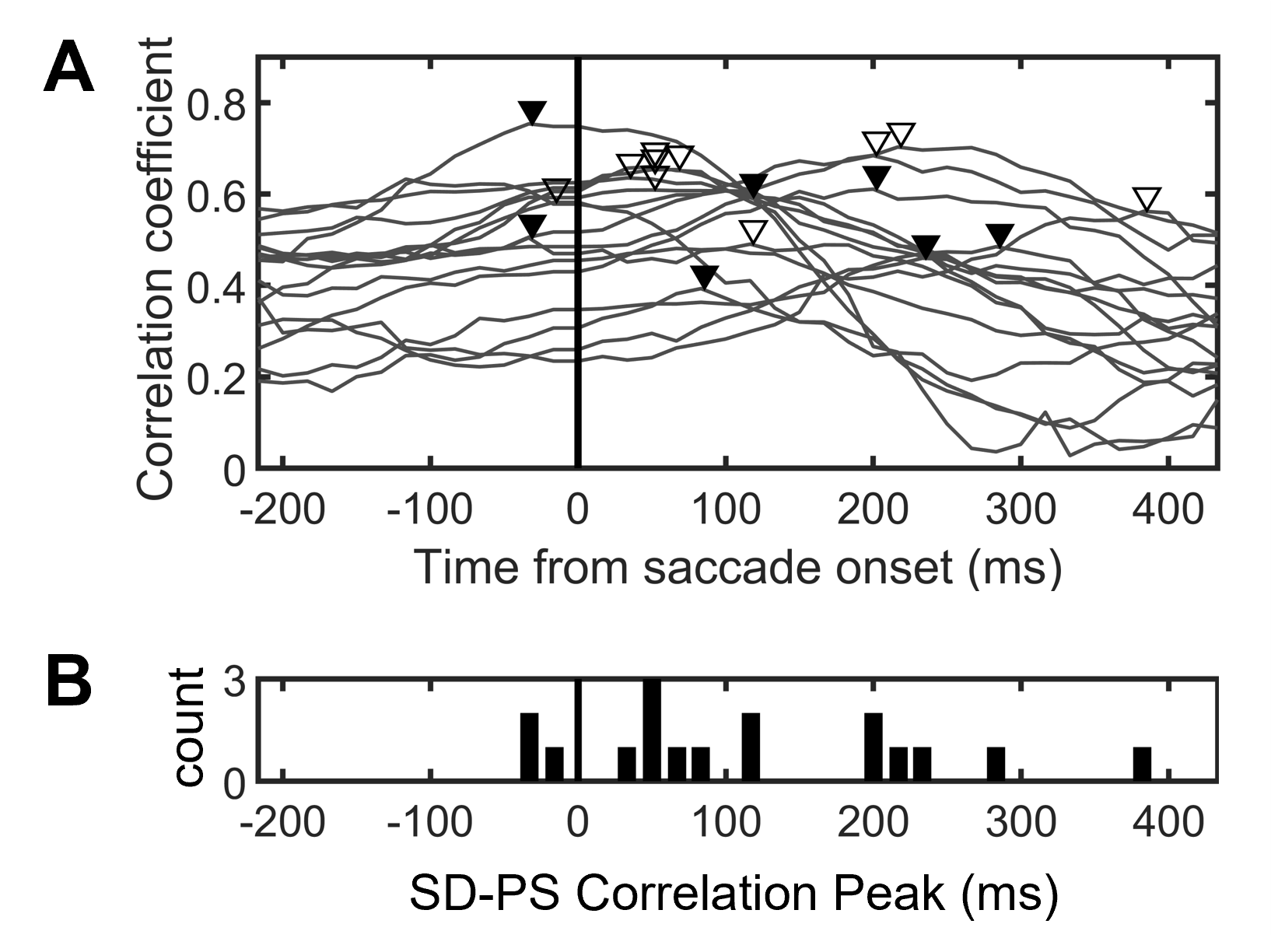} 
\end{center}
\justify 
\color{Gray}
\hypertarget{Figure 4}{\textbf{Figure 4. Distribution of SD–PS peak times across participants.}}
\textbf{A.} Cross correlation profiles for all 17 participants. Each line represents the correlation between SD and PS as a function of time lag. Inverted triangles indicate the SD–PS peak time (black: calligraphers; white: non calligraphers). The vertical line at t = 0 marks saccade onset. \textbf{B.} Histogram of SD–PS peak times. The distribution shows substantial individual variability, with peak times ranging from approximately -50 ms to +400 ms.

\newpage

\begin{center}
\includegraphics[width=0.72\linewidth]{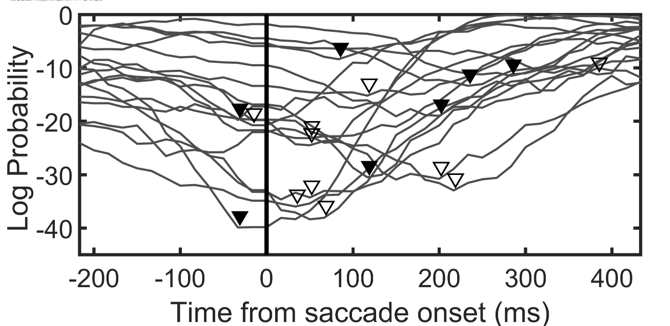} 
\end{center}
\justify 
\color{Gray}
\hypertarget{Figure 5}{\textbf{Figure 5. Statistical significance of SD–PS correlations across time lags.}}
\newline
Statistical significance profiles for all 17 participants. The vertical axis represents the log probability of the correlation between SD and PS. The horizontal axis shows the time lag relative to saccade onset (t = 0), consistent with \textbf{Figure 4}. Each line represents an individual participant, with inverted triangles indicating the SD–PS peak time (black: calligraphers; white: non calligraphers). Most participants show a sharp increase in significance (drop in log probability) at their respective peak times, confirming the robustness of the temporal coupling.

\begin{center}
\includegraphics[width=0.4\linewidth]{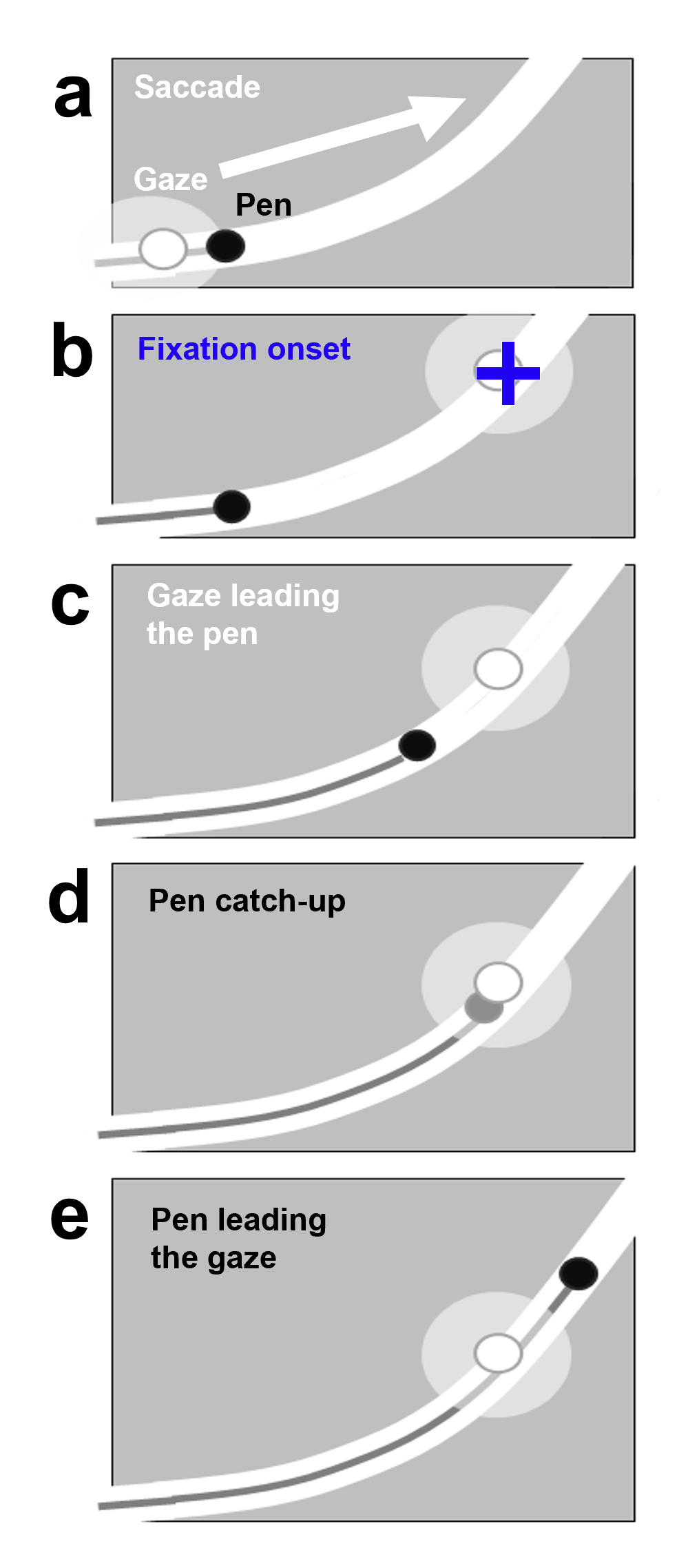} 
\end{center}
\justify 
\color{Gray}
\hypertarget{Figure 6}{\textbf{Figure 6. Schematic illustration of gaze–hand coordination patterns.}}
\newline
Representative time series showing the relative positions of gaze (open circle) and pen (black circle) on the target line (white line). \textbf{a.} Timing of saccade onset (arrow indicates saccade direction). \textbf{b.} Initiation of fixation following the saccade. \textbf{c.} Phase in which the gaze maintains a spatial lead ahead of the pen. \textbf{d.} Time point at which the pen reduces the distance to the gaze position and catches up. \textbf{e.} Phase in which the pen briefly precedes the gaze position. This sequence repeats throughout the tracing task.

\begin{center}
\includegraphics[width=0.72\linewidth]{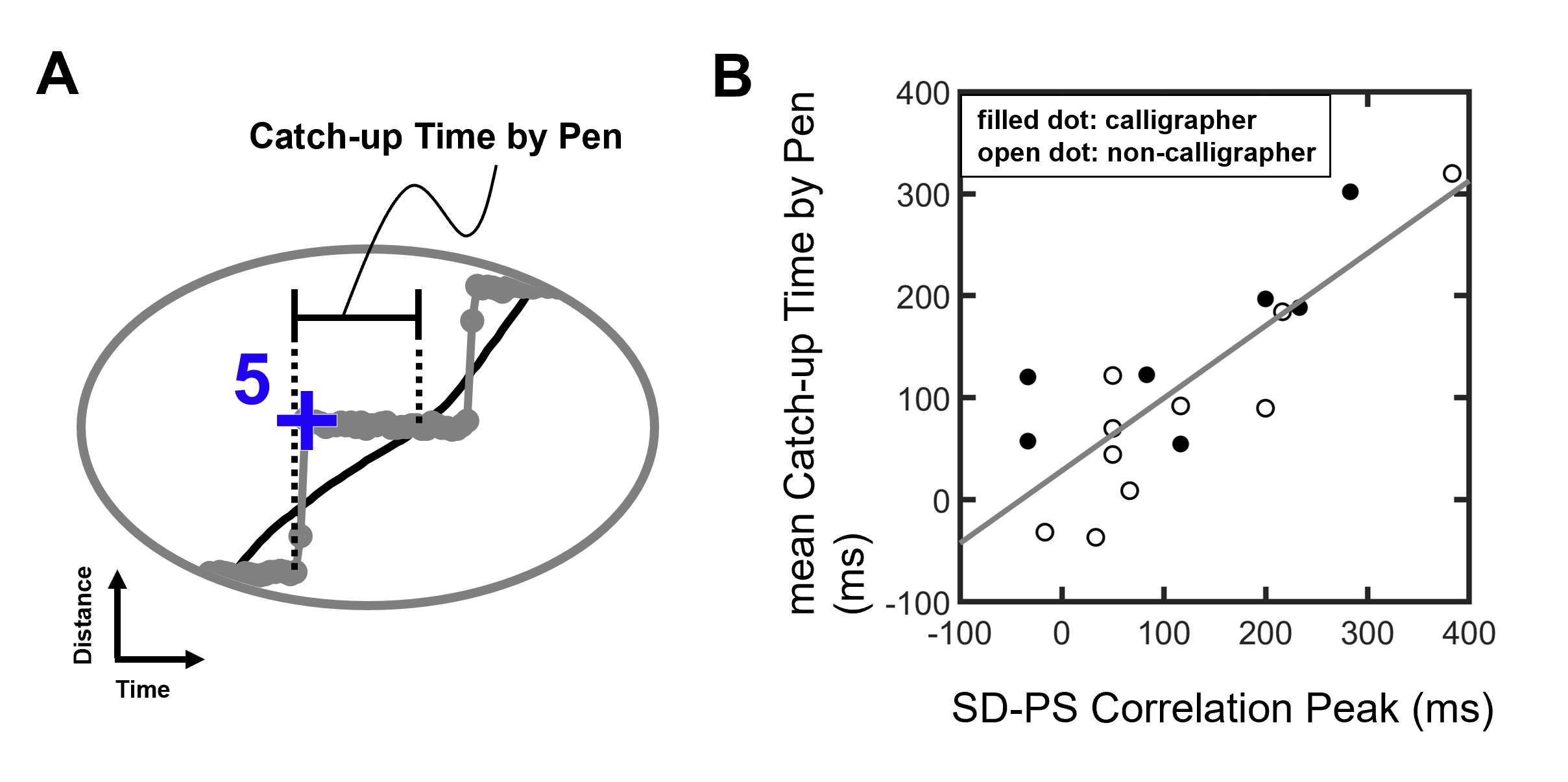} 
\end{center}
\justify 
\color{Gray}
\hypertarget{Figure 7}{\textbf{Figure 7. Relationship between the SD–PS peak time and the pen’s catch up timing.}}
\newline
\textbf{A.} Schematic definition of the pen catch up time, illustrating the interval between a saccadic gaze shift and the moment the pen reaches the same spatial position. \textbf{B.} Correlation between SD–PS peak time and mean catch up time across participants (n = 17). Filled dots represent calligraphers; open dots represent non calligraphers. A strong positive correlation indicates that the predictive window is tightly coupled with the spatial dynamics of the hand.
\bigskip\newline

\begin{center}
\includegraphics[width=0.8\linewidth]{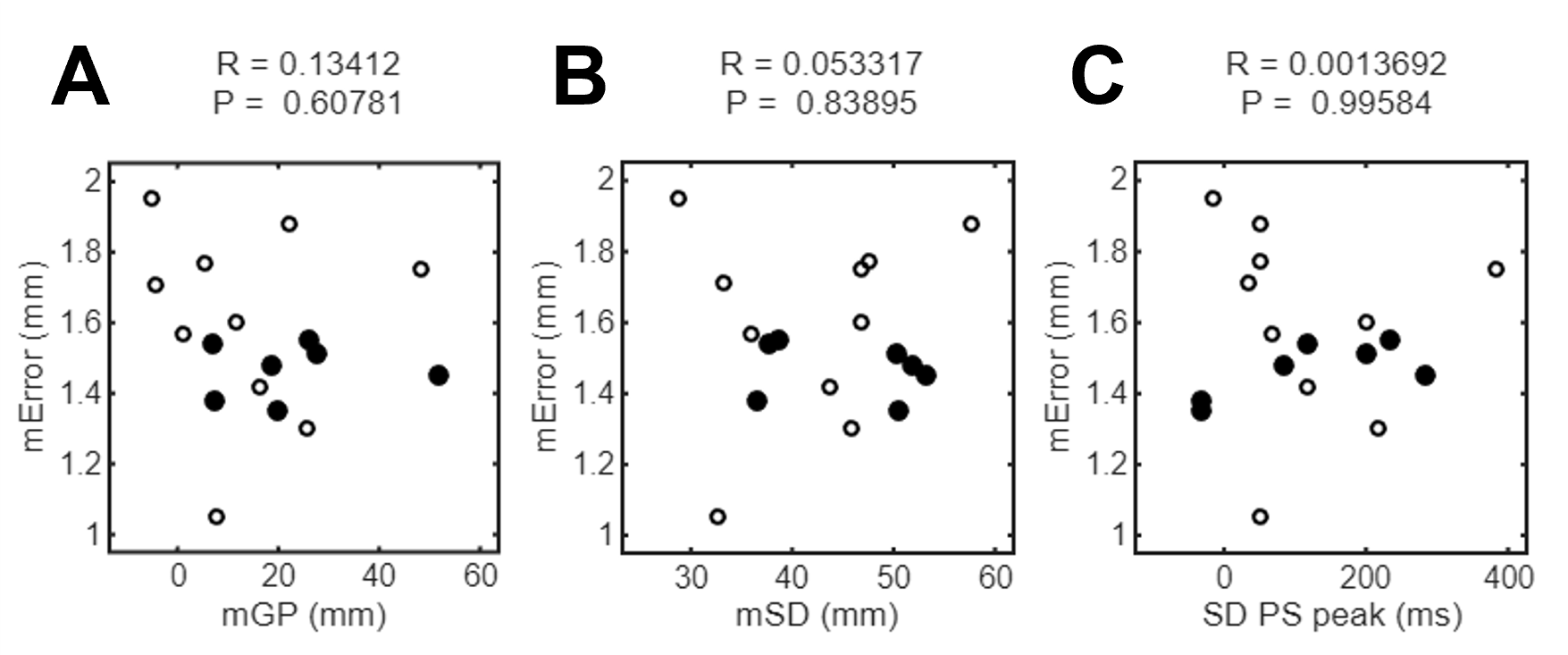} 
\end{center}
\justify 
\color{Gray}
\hypertarget{Figure 8}{\textbf{Figure 8. Correlation between oculomotor parameters and tracing accuracy.}}
\newline
\textbf{A–C.} Scatter plots showing the relationship between mean tracing error (mError) and three gaze related indices: \textbf{A.} mGP vs. mError. \textbf{B.} mSD vs. mError. \textbf{C.} SD–PS peak time vs. mError. Each dot represents an individual participant (filled: calligraphers; open: non calligraphers). Correlation coefficients (R) and P values are shown above each plot. The absence of significant correlations indicates that individual Predictive Protocols do not directly determine tracing accuracy.
\end{adjustwidth}

\includepdf[pages=-]{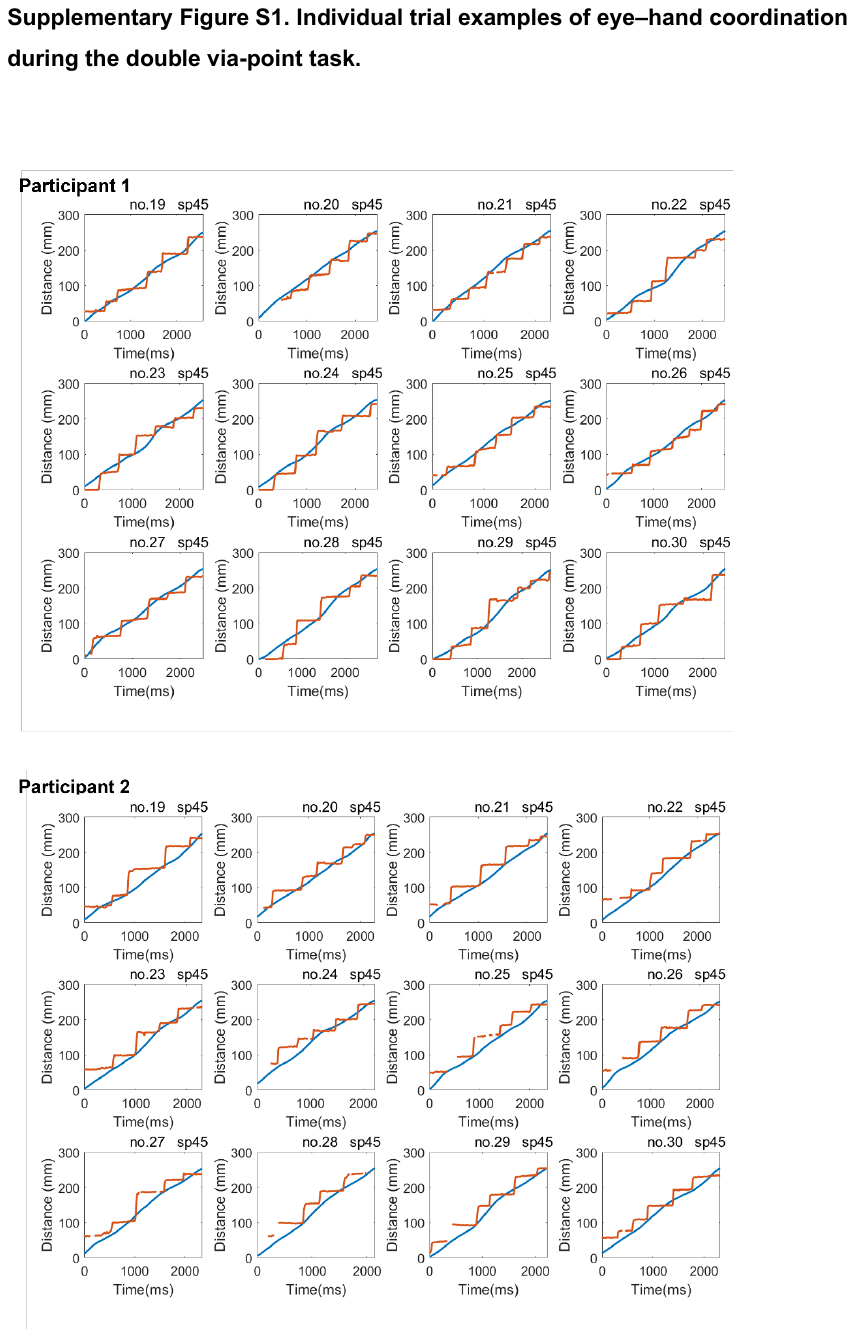}

\end{document}